# Ultracompact single-photon sources of linearly polarized vortex beams

Xujing Liu[1,2], Yinhui Kan[2*], Shailesh Kumar[2], Liudmilla F. Kulikova[3], Valery A. Davydov[3], Viatcheslav N. Agafonov[4], Changying Zhao[1*], Sergey I. Bozhevolnyi[2*]

[1]Institute of Engineering Thermophysics, Shanghai Jiao Tong University, Shanghai, 200240, China.

[2]Centre for Nano Optics, University of Southern Denmark, DK-5230 Odense M, Denmark.

[3]L.F. Vereshchagin Institute for High Pressure Physics, Russian Academy of Sciences, Troitsk, Moscow, 142190, Russia.

[4]GREMAN, CNRS, UMR 7347, INSA CVL, Université de Tours, 37200 Tours, France.

*Corresponding author. Email: yk@mci.sdu.dk; changying.zhao@sjtu.edu.cn; seib@mci.sdu.dk

**Ultracompact chip-integrated single-photon sources of collimated beams with polarization-encoded states are crucial for integrated quantum technologies. However, most of currently available single-photon sources rely on external bulky optical components to shape the polarization and phase front of emitted photon beams. Efficient integration of quantum emitters with beam shaping and polarization encoding functionalities remains so far elusive. Here, we present ultracompact single-photon sources of linearly polarized vortex beams based on chip-integrated quantum emitter-coupled metasurfaces, which are meticulously designed by fully exploiting the potential of nanobrick arrayed metasurfaces. We first demonstrate on-chip single-photon generation of high-purity linearly polarized vortex beams with prescribed topological charges of -1, 0, and +1. We further realize multiplexing of single-photon emission channels with orthogonal linear polarizations carrying different topological charges and demonstrate their entanglement. Our work illustrates the potential and feasibility of ultracompact quantum emitter-coupled metasurfaces as a new quantum optics platform for realizing chip-integrated high-dimensional single-photon sources.**

Generation of photon beams with specific polarizations and phase fronts is at the cornerstone of both classical and quantum optics[1,2]. Integration of both classical and quantum photon sources with beam shaping and polarization encoding functionalities has always been on the agenda and recently been forced to radically speed up because of rapidly growing demands of integration and miniaturization in photonics[3,4]. Furthermore, conventional approaches for molding single-photon emission, relying on external bulky devices, such as polarization filters and spiral phase plates, decrease not only the compactness of single-photon configurations but also their efficiency and brightness. Recently developed planar optical components, metasurfaces[5-7], have been introduced to manipulate incident single-photon beams[8], although been so far utilized as external components[9,10], similarly to their bulk counterparts. Therefore, it is still an ongoing challenge to realize ultracompact chip-integrated single-photon sources of collimated beams encoded with desired polarization states required for highly integrated quantum systems[11,12].

Engineering the immediate environment of quantum emitters (QEs) provides a promising way for realization of compact single-photon sources with tailored emission characteristics[13-15]. It has been demonstrated that utilizing appropriately designed nanocavities and nanoantennas can significantly increase photon emission rates[16-18], realize polarized[19,20] and directional[21,22] emission. These developments suggest a potential route for at-source solving the intrinsic problems of free-standing QEs, such as low efficiency, poor polarization, and omnidirectional emission[23-25]. Recently, spin-orbit coupled nano-configurations have been designed and integrated with QEs, featuring simultaneous generation of right and left circular polarized modes[26-28]. Although this research demonstrated interesting approaches for on-chip generation of single photons carrying orbital angular momentums (OAMs), these suffer from the *fundamental limitation* resulting in an *inevitable* generation of both circular polarizations having the same OAM difference[29]. Up to now, on-chip generation of single-photon beams with high-purity linearly polarized vortices has not been realized. Moreover, the design approach that would enable on-chip QE-coupled metasurfaces to accurately control the polarization and phase of emitted photons has so far been elusive.

Here, we demonstrate ultracompact chip-integrated single-photon sources of linearly polarized vortex beams based on QE-coupled metasurfaces. A general approach for designing QE-coupled nanobrick-arrayed metasurfaces is developed, facilitating the control of the polarization, OAM, and entanglement of multichannel single-photon emission. This breakthrough development is based on fully exploiting the potential of nanobrick arrayed metasurfaces with all three metasurface aspects, nanobrick dimensions with its local orientation and global positioning, being fine-tuned in concert. Our approach is thus in stark contrast with recently presented spin-orbit nano-configurations[26-28] featuring limited freedom in both design parameters and operation characteristics. We first demonstrate single-photon generation of high-purity linearly polarized (LP) vortex beams with prescribed different OAMs by using meticulously designed nanobrick-arrayed metasurfaces for outcoupling surface plasmon polariton (SPP) modes that are efficiently excited by a QE. Remarkably, the design strategy enables *simultaneous* generation of high-purity *orthogonal* LP single-photon beams with *arbitrary* different OAMs, meeting thereby growing demands for increasing the quantum information channel capacity. Thus, by selectively covering different azimuthal metasurface areas with differently arranged nanobricks, we realize multiplexing of single-photon emission channels with orthogonal LPs carrying different topological charges (OAMs) and confirm their entanglement.

**Operation principle**

The proposed single-photon sources represent chip-integrated QE-coupled metasurfaces, consisting of dielectric anisotropic nanobricks designed around preselected QEs generating upon excitation radially diverging SPP waves supported by silica-covered silver (Ag) substrates. Single photon LP emission (Fig. 1a), single photon LP-OAM (Fig. 1b), and entangled LP-OAMs emission (Fig. 1c) are realized by three configurations with full exploitation of the three aspects of metasurfaces: element, trajectory, and arrangement of area. Nanodiamonds containing single germanium vacancy centre (GeV-ND) are used as QEs atop silica ($SiO_2$) spacer covered silver (Ag) substrate. GeV-ND, excited by a 532 nm radially polarized pump laser, features vertical electric dipole that excites radial SPPs supported by the air-$SiO_2$-Ag interface (Fig. 1d). The anisotropic nanobrick, made of hydrogen silsesquioxane (HSQ), is designed with a height of 150 nm, a width of 100 nm, and a length of 350 nm. Using an array of the designed nanobricks with the period of 510 nm along the radial direction, QE-excited SPPs can be efficiently coupled into free space emission at 602 nm that coincides with emission peak of GeV-ND. Figure 1e plots the simulated LP performance of outcoupled light from array of nanobricks (area $7 \times 8\ \mu m^2$) as a function of orientation angle ($\varphi$) and azimuthal direction ($\theta$) of propagating SPPs. The LP performance is evaluated by the Stokes parameter $S_1$, which is the normalized intensity difference of two orthogonal linear polarization states along the *x*- and *y*-axis (LPx and LPy), denoted as $S_1 = (I_x - I_y)/(I_x + I_y)$. For small $\theta$ (e.g., $\theta = 0°$, SPP propagating along *x* direction), $S_1$ maintains high value for a broad range of orientation angles ($S_1$ attains up to 0.998), which means LPx dominates the outcoupled photon emission no matter what orientation nanobricks is. For large $\theta$ (e.g., $30° < \theta < 60°$), only specific orientation can make the LPx contribution larger than LPy. When $\theta > 60°$, LPy will be always larger than LPx no matter what $\varphi$ is. It should be noted that LPx and LPy would have equal contribution (i.e., $I_x = I_y$) for $\theta = 45°$ when $\varphi = 45°$, whereas $I_x$ can be larger than $I_y$ for $\theta = 45°$ by carefully setting $\varphi$, e.g., $\varphi = 0°$ leading to $I_x/I_y = 4.4$. The theoretical prediction of $\varphi$ and $\theta$ satisfies $\varphi \cong \theta - 45°$ for optimum $I_x/I_y$, coinciding well with the simulation (Supplementary Note. S1, Figs. 1-3). Therefore, to generate high purity LPx, we azimuthally arrange nanobricks around QE with $\varphi = \theta - 45°$ and the array is periodic in radial direction. We do not put any nanobricks in the areas of $60° < \theta \leq 120°$ and $240° < \theta \leq 300°$ as these areas mainly contribute to LPy (Supplementary Fig. 4). Besides orientation, the distance of nanobrick to QE can control phase shifts of SPPs, which provides another degree of freedom to achieve high-purity LP single photon emission, and further realizing LP carrying OAM (LP-OAM). The start radii of two parts (on the left and right of QE) have a difference of $\lambda_{spp}/2$, enabling an initial phase difference of $\pi$ for generation of unidirectional LP photons, where $\lambda_{spp}$ is the wavelength of SPP. Furthermore, arranging two sides following the Archimedean spiral trajectories with different start radii can provide an azimuthal phase shift (in addition to initial phase difference) and realize LP-OAM with varying phase accumulations of characteristic $2\ell\pi$ ($\ell = 0, \pm 1, \pm 2 ...$). Taking advantage of the generating LP-OAM, the design can be further extended to entangled LP-OAM by dividing different operation areas and combining the two orthogonal LP states encoding different helical phase fronts carrying distinctive OAMs.

**GeV-ND single-photon source**

In this work, we choose GeV-ND as QE since it can generate single photons at room temperature with a narrow emission peak at 602 nm[30]. GeV-NDs with the size around 100 nm are dispersed randomly on $SiO_2$ spacer covered Ag substrate by spin-coating solutions containing GeV-NDs (Supplementary Fig. 5 and 6). Before fabricating metasurfaces around them, GeV-NDs are characterized by a series of measurements to preselect those with large vertical transition component and containing single GeV centres according to fluorescence scan images, spectra, lifetimes, and auto-correlation relations (setup presented in Supplementary Fig. 7). GeV-NDs are excited using a tightly focused radial polarized laser beam (532 nm), which produces a strong electric field component ($E_z$) perpendicular to the surface for exciting vertical transition component of QEs. As depicted in Fig. 2a, those GeV-NDs whose fluorescence scanning image show doughnut-shape patterns, indicating the vertical electric dipole features, are selected as QEs to efficiently excite radial SPPs at the air-$SiO_2$-Ag interface. After NDs being positioned with developed nanoscale alignment technique[25], metasurfaces are precisely fabricated around NDs with standard electron-beam lithography (EBL) (see details in Methods, Supplementary Fig. 5 and 6). Figure 2b shows an atomic force microscope (AFM) image of the fabricated device for LP single-photon emission, where the GeV-ND is observed located in the centre. The metasurface with a 150 nm thickness shows ultra-compactness of the proposed single photon sources. The spectra of the same GeV-ND before and after being coupled with metasurface show narrow-band fluorescence emission peaks around 602 nm, where the counts increase after coupling with metasurface. Correspondingly, the lifetimes of GeV-ND decreases from $\tau = 12.1$ ns to $\tau = 7.6$ ns due to Purcell enhancement. The second-order correlation functions are measured by Hanbury Brown-Twiss setup, showing $g^{(2)}(0) = 0.17$ and 0.23 before and after coupling with metasurface, respectively (Fig. 4d), indicating excellent and stable single-photon properties at room temperature.

**LP-OAM generation**

We demonstrate three single-photon sources of LP-OAM by arranging nanobricks with different trajectories for topological charges of $\ell = 0, +1,$ and $-1$, as illustrated in Fig. 3a from top to bottom, respectively. In all cases, the start radii of trajectories of the right side ($r_1$) and left side ($r_2$) are: $r_1 = 1.5\lambda_{spp}$ and $r_2 = 2\lambda_{spp}$ where a difference of $\lambda_{spp}/2$ provides an initial phase difference of $\pi$ for the SPPs propagating on the left and right direction before coupling with nanobricks, enabling thereby the generation of unidirectional LP photon emission contributed from both sides. For topological charge of $\ell = 0$, the trajectory is concentric at each side. The bricks in radial direction are periodically distributed with the same orientation. The angle of distributing nanobricks is optimized as $\alpha = 60°$, considering the tradeoff between LP performance and the outcoupling efficiency (Supplementary Fig. 4). To realize the OAM state, nanobricks at two sides are arranged with azimuthally varied distances from GeV-ND, following two sperate counterclockwise Archimedean spiral trajectories with different start radii ($r_1$, $r_2$). The radius of the spiral trajectory changes $\lambda_{spp}$ from head to tail, leading to a phase accumulation of $2\pi$. The in-plane propagating SPPs interacting with the metasurface and subsequently being coupled to far-field emission will carry spiral phase ($\ell = +1$). Similarly, the clockwise Archimedean spiral trajectory accumulates phase from 0 to $-2\pi$, resulting in photon emission with OAM of $\ell = -1$.

The scanning electron microscope (SEM) images of the fabricated GeV-ND coupled metasurfaces with different trajectories are shown in Fig. 3b. The Stokes parameters $S_1$ are directly measured with the photon emission passing through a linear polarizer. Figure 3c presents the superimposition of measured emission pattern and $S_1$ with numerical aperture (NA) of 0.9. The insets are comparisons between simulated and measured far-field emission (NA = 0.2). For $\ell = 0$, the emission forms a collimated spot (with the divergence angle of 2.5°) dominated by near-perfect LPx (with the peak of $S_1 = 0.99$). By integrating the decomposed component within divergence angle, the ratio of $I_x/I_y$ attains 40, demonstrating high-purity LP single photon emission. For $\ell = \pm 1$, the emission shows doughnut-shape patterns with LP ($S_1 = 0.97$ at peak intensities). The corresponding photon emission carry LP-OAM, where OAM leads to the doughnut intensity distributions in the momentum space.

To characterize the OAM states, the topological charge of emission is measured by projecting the far-field emission to different spiral phase distributions of spatial light modulator (SLM), shown in each top row of Fig. 3d (setup presented in Supplementary Fig. 8). The projected far-field emission pattern forms a spot when the topological charge is opposite with the SLM phase profile. As shown in Fig. 3d, it ascertains generation of the LP-OAM beams carrying the corresponding topological charges $\ell = 0$, +1, and −1 with different trajectories. The simulated and measured emission patterns after being projected with other SLM phase profiles are shown in Supplementary Figs. 9 and 10. It is worth noting that this design principle can be applied to various QEs with different wavelengths by changing the period and element size of metasurface (Supplementary Figs. 11 and 12). Experimental performance of high-purity LP ($\ell = 0$) photon sources for nanodiamonds containing nitrogen vacancy centres (with peak emission at 670 nm) has also been demonstrated (Supplementary Fig. 12).

**Entangled LP-OAM**

Taking advantage of LP-OAM superposition state in single channel, it is possible to realize multi-dimensional LP single photon emission. Here we show the generation of two-channel LP-OAM photon emission by taking full use of azimuthal area (Fig. 4a). Two types of metasurfaces, separated by angles $\alpha$ and $\beta$, are used to generate two separate channels for orthogonal LPs (i.e., LPx and LPy), where green part is responsible for LPx channel carrying OAM states of $\ell_x = 0$, and the purple part is responsible for LPy channel carrying OAM states of $\ell_y = -1$. The insets illustrate the simulated far-field emission patterns and corresponding phase profiles of two LP channels. Figure 4b presents the SEM image of the fabricated device. LPx and LPy channels are measured by rotating the polarizer, showing a Gaussian beam spot and a doughnut-shape (Fig. 4c), respectively, which corresponds well with the theoretical prediction (measurement of topological charge can be seen in Fig. S13). The output entangled photon state can be written as $|\psi\rangle = a_x|X\rangle|\ell_x = 0\rangle + a_y e^{i\delta}|Y\rangle|\ell_y = -1\rangle)$, where $\delta$ is the phase between two entangled states, $a_x$ and $a_y$ are real positive numbers with $a_x^2 + a_y^2 = 1$, as required for normalization. To validate the entanglement between LPx and LPy channels and characterize the entangled states, quantum state tomography is performed with 16 projection measurements with different combination of polarization and OAM basis[31,32]. The simulated (Fig. 4d) and measured (Fig. 4e) density matrices with real and imaginary parts show the entangled states. The fidelity between the reconstructed density matrices of the predicted and experimentally obtained states is defined as $F(\rho_t, \rho_{exp}) = \left[\text{Tr}\left(\sqrt{\sqrt{\rho_t}\rho_{exp}\sqrt{\rho_t}}\right)\right]^2$, where $\rho_t$ and $\rho_{exp}$ are the density matrices of the entangled state of predicted results and the measured results, respectively. The fidelity is 0.78, demonstrating the entangled LP-OAM of two channels. It is worth noting that the emission percentage of two channel as well as the quantum entanglement are tunable by varying the angle of two part of metasurfaces (Supplementary Fig. 14), providing a promising way capable of producing two different OAM states and two different polarization states in a reconfigurable manner. In this context, the proposed on-chip QE-coupled metasurfaces may be extended further as a platform to provide spatial channels for advanced information modulation[33], high-dimensional quantum communication[34], and quantum detection[35].

**Conclusion**

We propose and implement a series of ultracompact single-photon sources based on chip-integrated QE-coupled metasurfaces enabling the generation of versatile well-collimated LP-OAM single-photon beams at room temperature. Encoding LP photons with OAM states offers new degrees of freedom meeting thereby ever-increasing demands for increasing the quantum information channel capacity. We anticipate that the ultracompact QE-coupled metasurfaces will provide new pathways for further development of on-chip integrated quantum photonic systems, such as, for example, realizing compact quantum information processing[36]. The presented design strategy makes full use of the available surface area and arrayed nanobrick configuration, including nanobrick dimensions with its local orientation and global positioning, which can be viewed as another advance of on-chip QE-coupled metasurfaces compared with the previously reported structures[26-28]. With this toolbox at hand, one is well equipped to generate very complex structured single-photon beams in multiple channels, which holds the promise for future development of quantum channel division/multiplexing circuits for high-density data and multi-dimensional information processing [34, 37].

**Methods**

**Numerical simulations.** Numerical simulations of QE-coupled metasurfaces were performed with three-dimensional (3D) finite-difference time-domain (FDTD) method. The quantum emitter was simulated as an *z*-direction electric dipole radiating at 602 nm, fitting the emission peak of GeV-ND. The dipole was placed 30 nm atop the $SiO_2$ spacer (20 nm) and silver film (120 nm) and in the centre of dielectric nanostructures (with thickness of 150 nm, refractive index 1.41). The periodicity of the metasurfaces was set be equal to the SPP wavelength ($\lambda_{spp} = 510$ nm with the effective SPP mode index $n_{spp} = 1.18$). A two-dimensional monitor 30 nm atop the configuration was employed to collect far-filed electric field by near- to far-field transformation method.

**Device fabrication.** Fabrication of QE-coupled metasurfaces followed a well-established sequence of technological steps (Supplementary Fig. 5). A 150 nm thick silver-film was first deposited on the silicon substrate by the thermal evaporation, followed by a 20-nm-thick $SiO_2$ layer being deposited by the magnetron sputtering method. Then a group of gold markers were fabricated on the substrate by a combined process of EBL (JEOL-6490 system, accelerating voltage 30 kV), gold deposition, and lift-off. Subsequently, solution of GeV-NDs [38] were spin-coated with proper diluted concentration. After preselecting GeV-NDs by the fluorescence scan and characterizing quantum properties, the positions of candidates are determined by the dark-field microscope image with the precise alignment procedure (Supplementary Fig. 6), a negative photoresist (HSQ) was subsequently spin-coated at 4000 rpm, for 45s, and heated at the hotplate on 160 °C for 2 minutes to form a ~150 nm layer, which was examined by the AFM (atomic force microscope, NT-MDT NTEGRA). Subsequently, a second round EBL was then carried out on the resist around GeV-NDs. Finally, QE-coupled metasurfaces were obtained by development with the tetramethylammonium hydroxide solution (for 4 min) and isopropanol (60 s).

**Optical characterization.** The QE-coupled metasurfaces was characterized with the experimental setup as shown in Supplementary Figs. 7 and 8. Radially polarized 532 nm laser beam was used to drive and preselect GeV-NDs that can form a dominant dipole moment perpendicular to the substrate surface. The excitation and collection of photon emission were by the same objective with NA = 0.9 (The Olympus MPLFLN ×100). Fluorescence imaging were performed by a synchronous movement of piezo-stage with mounted sample and projecting the collected fluorescence emission to the avalanche photo diodes (APDs). Decay-rate measurements was carried out with a pulsed laser (Pico Quant LDH-PFA-530L) at 1 MHz together with APD1 (τ-SPAD, PicoQuant). Fluorescence spectra were measured using the spectrometer (Andor Ultra 888 USB3-BV) operating within 540-820 nm. Second-order correlation was recorded by registering the temporal delay between photon detection events between APD1 and APD2 in a start-stop configuration, using an electronic timing box (Picharp-300; Pico quant). The Stokes parameters were measured to characterize the polarisation states. $s_1$ and $s_2$ were measured by orienting the linear polarizer and taking Fourier plane images. $s_3$ was measured by the combination of a quarter-wave plate and a linear polarizer (Supplementary Note 2). An SLM was used to generate the computed holograms with different topological charges. The reflected light from the incident light was filtered out by a set of dichroic mirrors (Semrock FF535-SDi01/FF552-Di02) and with a long-pass filter 550 nm (Thorlabs FELH0550) and a band-pass filter 605 nm ± 8 nm. The quantum state topography was performed based on the polarization and OAM coincidence with a set of superpositions of polarization and OAM basis (Supplementary Note 3).

**Acknowledgements**

The authors acknowledge the support from National Natural Science Foundation of China (Grant No. 62105150) (Y.H.K.) and (No. 52120105009) (C.Y.Z.), European Union's Horizon Europe research and innovation programme under the Marie Skłodowska-Curie Action (Grant agreement No. 101064471) (Y.H.K.), Natural Science Foundation of Jiangsu Province (BK20210289) (Y.H.K.), Shanghai Key Fundamental Research Grant (No. 20JC1414800) (C.Y.Z.), State Key Laboratory of Advanced Optical Communication Systems Networks of China (2022GZKF023) (Y.H.K.), and Villum Kann Rasmussen Foundation (Award in Technical and Natural Sciences 2019) (S.I.B.).

**Author contributions**

S.I.B. and Y.H.K. conceived the idea. X.J.L. and Y.H.K. performed theoretical modelling. X.J.L. fabricated samples. X.J.L. and Y.H.K. with assistance from S.K. performed experimental measurement. L.F.K., V.A.D., and V.N.A synthesized the GeV nanodiamonds. X.J.L., Y.H.K., S.K., and S.I.B. analyzed the data. S.I.B., Y.H.K., and C.Y.Z supervised the project. X.J.L., Y.H.K., S.K., and S.I.B. wrote the manuscript with contributions from all authors.

**Competing interests**

The authors declare no competing interests.

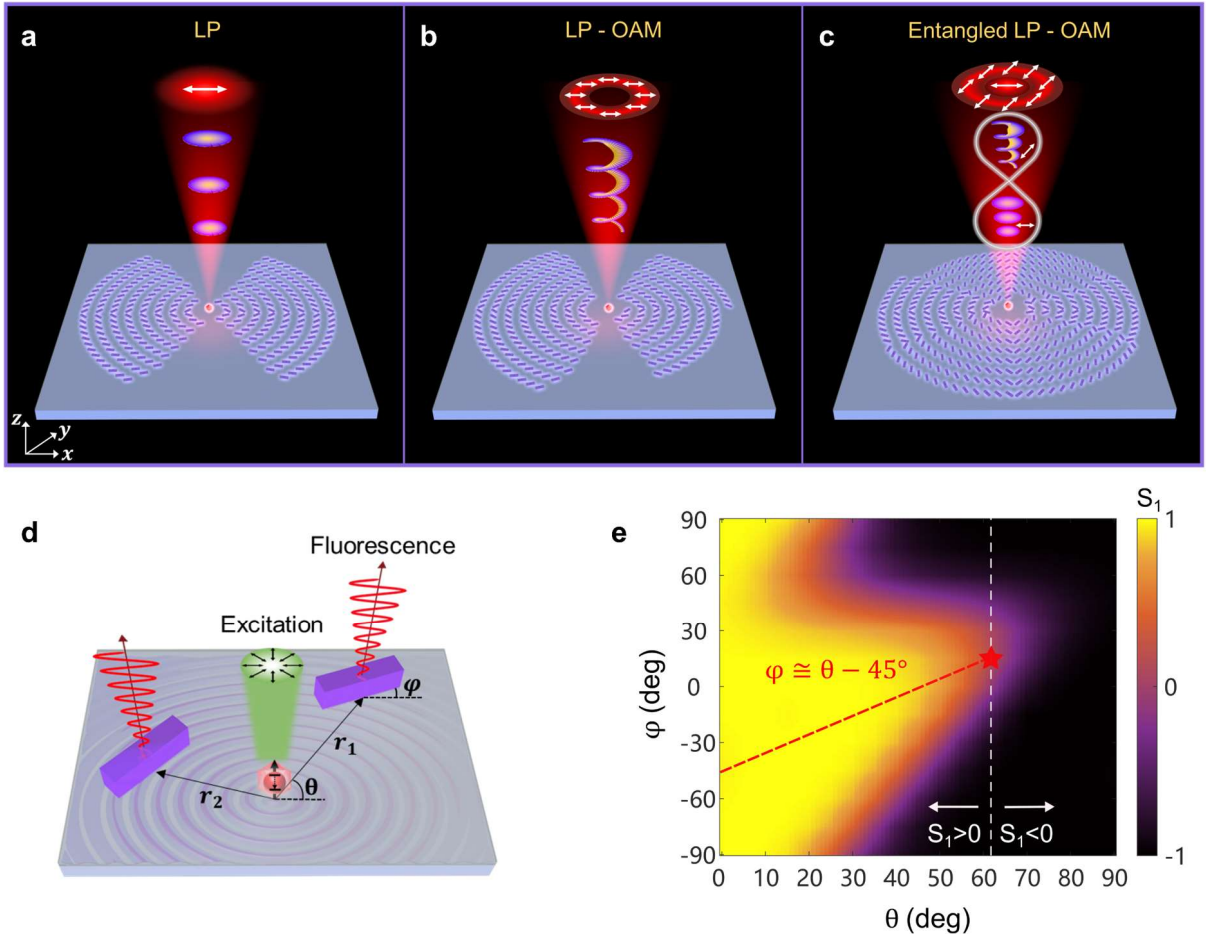

**Fig. 1 | Schematics of on-chip single-photon sources enabling linear polarization and OAM control. a**, Linearly polarized (along *x*-axis, LPx) single photon emission. The GeV-ND, used as QE, is on-chip surrounded by dielectric metasurfaces (with thickness of 150 nm) atop SiO$_2$ layer (20 nm) covered Ag substrate (120 nm). **b**, Linearly polarized (LPx) single photon emission carrying OAM ($\ell = 1$). **c**, Entangled linearly polarized (LPx and LPy) single photon emission carrying OAM ($\ell_x = 0$ and $\ell_y = -1$). **d**, Schematic of the coupling process of nanobricks and QEs. GeV-ND is excited by a 532 nm radial polarized pump laser, featuring vertical electric dipole and then exciting radially propagating SPPs along the surface, which are coupled out by nanobricks into photon emission. **e**, Simulated and theoretical guidance for brick arrangement. The color presents the simulated Stokes parameter $S_1$ of outcoupled photon emission from array of bricks with varying brick orientation ($\varphi$) and SPP propagating direction ($\theta$). The dashed line plots the theoretical relation of $\varphi$ and $\theta$ to obtain high LPx photon emission, the red star marks the terminal point of $\theta = 60°$ with $S_1 \cong 0$.

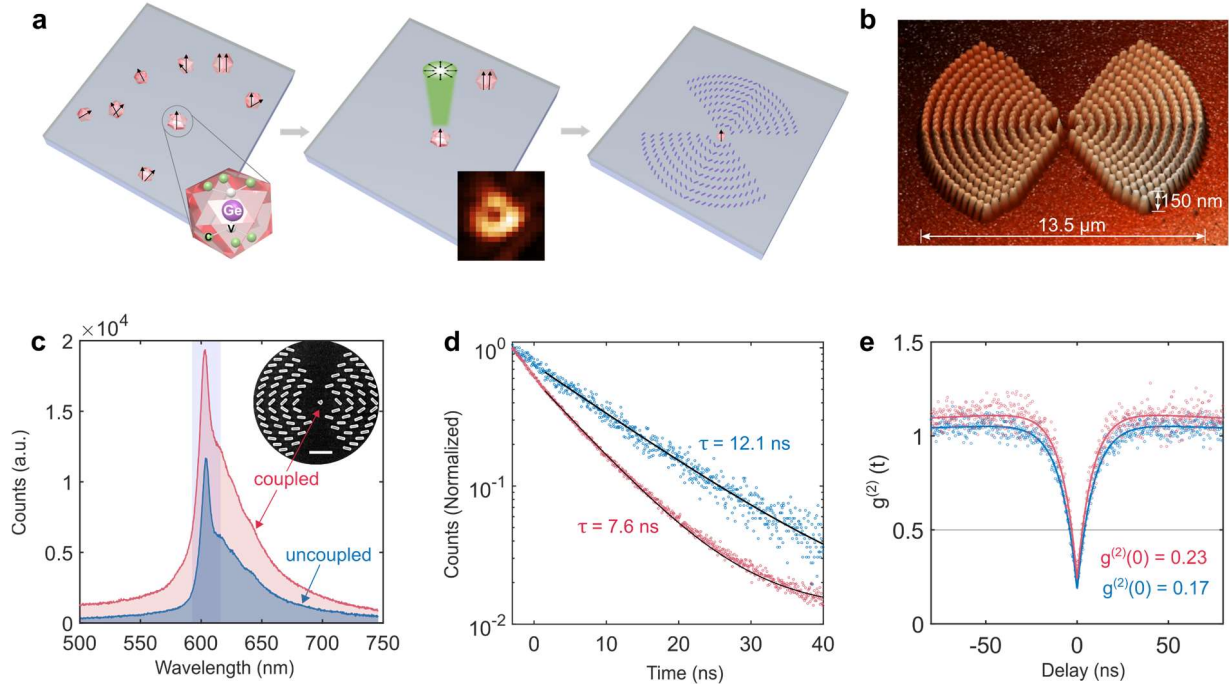

**Fig. 2 | Characterization of GeV-ND single-photon source before and after coupling with metasurfaces. a**, GeV-ND single photon source selection process. The left inset shows the schematic of a GeV-ND. The middle inset shows the fluorescence image (1.5 × 1.5 μm$^2$) of the selected GeV-ND, the doughnut shape indicates the GeV-ND with a vertical electric dipole feature when excited by the radially polarized pump laser. **b**, AFM image of the fabricated on-chip QE-coupled metasurface. **c**, Fluorescence spectra of GeV-ND uncoupled and coupled with metasurface. Scale bar of inset SEM image, 1 μm. **d**, Lifetime and **e**, Auto-correlation comparison of GeV-ND before and after coupling with on-chip metasurface.

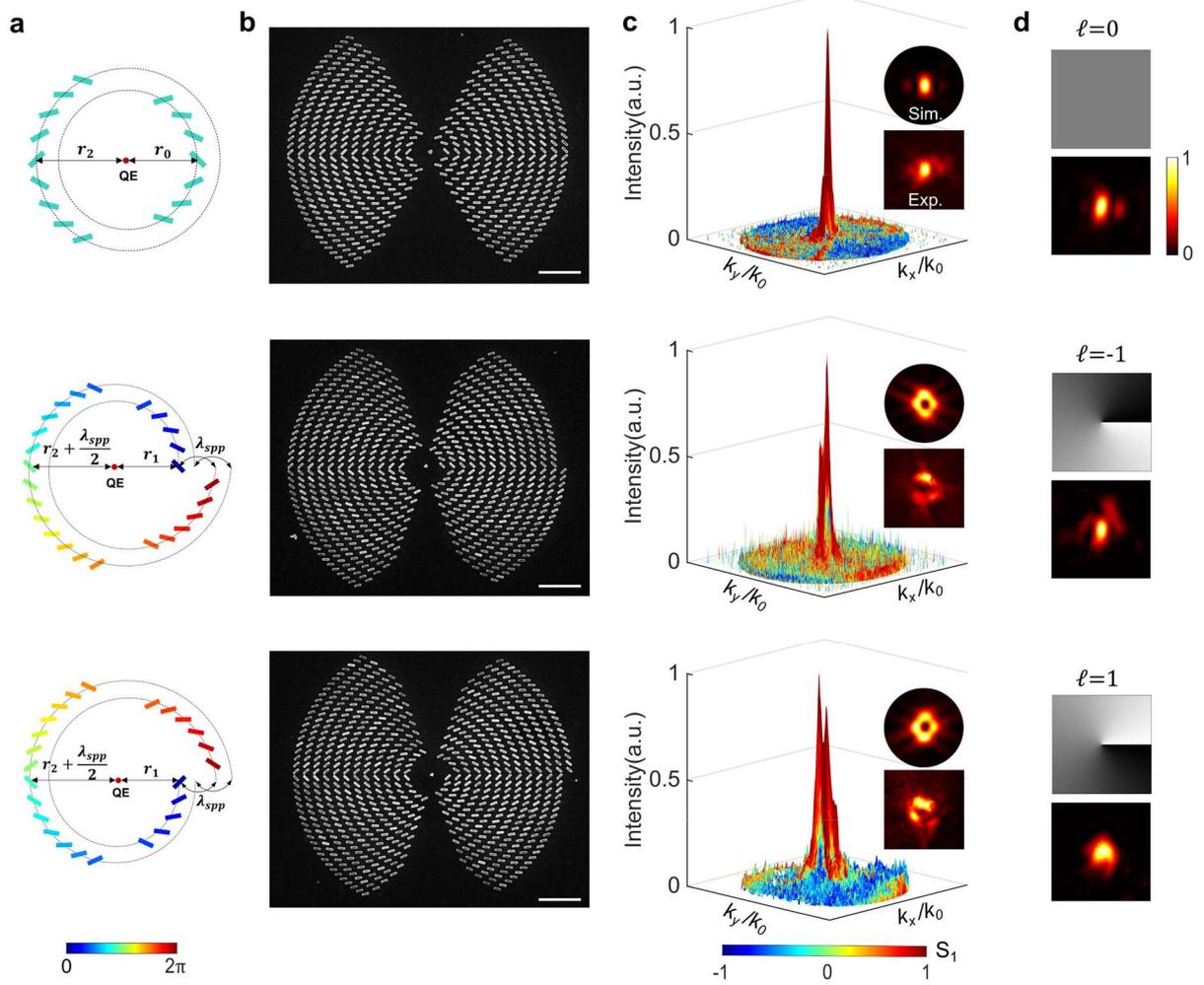

**Fig. 3 | Design and characterization of one-channel LP single photon source with vortices. a**, Azimuthal phase distribution of the designed QE-coupled metasurface (first winding). The dotted line shows the trajectories for azimuthally arranging the bricks around QE. **b**, SEM images of the fabricated QE-coupled metasurface. Scale bars, 2 μm. **c**, Superimposition of measured emission pattern and Stokes parameter $S_1$. The insets present the simulated (top) and measured (bottom) far-field emission patterns (NA = 0.2). **d**, Measured intensity patterns of LPx component after being projected to phase profile of SLM with topological charges of $\ell = 0$, $+1$, and $-1$ from top to bottom.

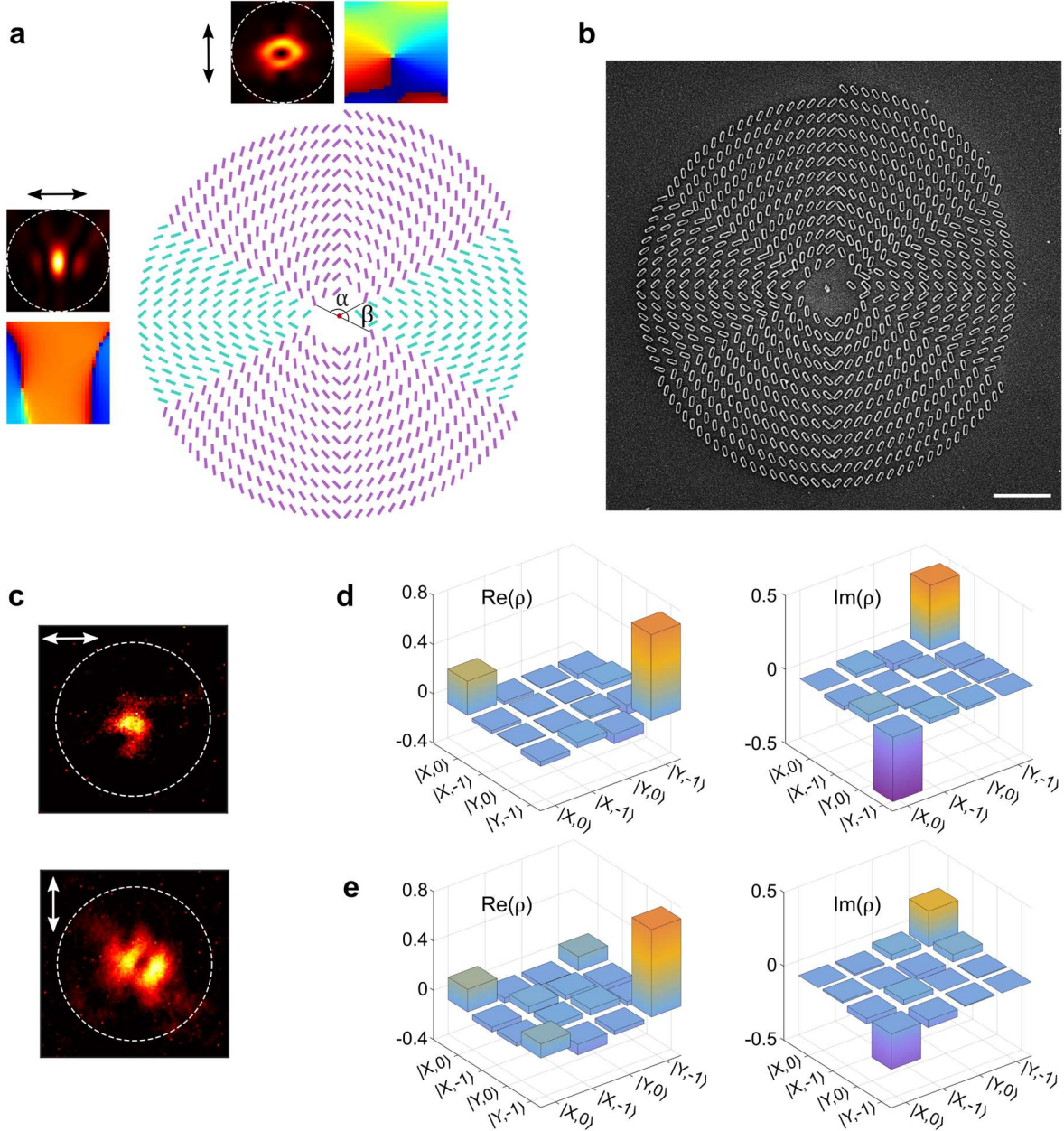

**Fig. 4. Demonstration of two-channel LP single photon source with vortices and entanglement**. **a**, Arrangement of chip-integrated QE-coupled metasurfaces for LPx and LPy channels with the topological charge of $\ell_x = 0$ and $\ell_y = -1$, respectively. The insets are the simulated emission patterns and phase distributions (NA = 0.2). **b**, SEM image of the fabricated two-channel QE-coupled metasurfaces. Scale bar, 2 μm. **c**, Measured emission patterns of LPx and LPy channels. The dashed line indicates the area of NA = 0.2. The arrow indicates the direction of polarization projection. **d, e,** The simulated (**d**) and measured (**e**) density matrices with real and imaginary parts.
17

# Supplementary information

---

## Ultracompact Single-Photon Sources with Linearly Polarized Vortices

Xujing Liu[1,2], Yinhui Kan[2]*, Shailesh Kumar[2], Liudmilla F. Kulikova[3], Valery A. Davydov[3], Viatcheslav N. Agafonov[4], Changying Zhao[1]*, Sergey I. Bozhevolnyi[2]*

1. Institute of Engineering Thermophysics, Shanghai Jiao Tong University, Shanghai, 200240, China.

2. Centre for Nano Optics, University of Southern Denmark, DK-5230 Odense M, Denmark.

3. L.F. Vereshchagin Institute for High Pressure Physics, Russian Academy of Sciences, Troitsk, Moscow, 142190, Russia

4. GREMAN, CNRS, UMR 7347, INSA CVL, Université de Tours, 37200 Tours, France

**This PDF file includes:**

Supplementary Note 1-3

Supplementary Figure 1-14

**Supplementary Note 1 | Theoretical analysis of SPP scattering by anisotropic nano-scatterers**

(1) SPP scattering by anisotropic nano-scatterers I

We consider the SPP incidence on a nano-scatterer of width $w$ and length $L$, inclined with respect to the SPP propagation direction as shown in Fig. S1a. We further assume that the scattered field can be decomposed into a component along nano-scatterer axis and perpendicular to it and that these component amplitudes are proportional to the nano-scatterer length and width, respectively, with the same proportionality coefficient (there is no phase considered and no resonance due to the dielectric nano-scatterers). All these results are in the following relations for the scattered field $x$- and $y$- components:

$$E_{scx} \sim E_x \cos\alpha\, L \cos\alpha + E_x \sin\alpha\, w \sin\alpha \sim (L-w)\cos^2\alpha + w \tag{S1}$$

$$E_{scy} \sim E_x \cos\alpha\, L \sin\alpha - E_x \sin\alpha\, w \cos\alpha \sim 0.5(L-w)\sin 2\alpha \tag{S2}$$

The ratio between these components can then be expressed as follows:

$$R = \frac{E_{scy}}{E_{scx}} = \frac{\sin 2\alpha}{2\left(\cos^2\alpha + \frac{w}{L-w}\right)} = \left\| a = \frac{w}{L} \right\| = \frac{\sin 2\alpha}{2\left(\cos^2\alpha + \frac{a}{1-a}\right)} \tag{S3}$$

where $a = w/L$. In the limit of vary narrow sticks, we have a very simple relation:

$$R = \frac{E_{scy}}{E_{scx}} \to \frac{\sin\alpha}{\cos\alpha} \tag{S4}$$

But this relation cannot be used for $\alpha = 90°$, because, in this limit $R = 0$ for obvious reasons. One should also take into account the scattering efficiency (the normalized intensity of the scattered-out field):

$$I_{scy} = \left(\frac{E_{scy}}{E_x}\right)^2 = 0.25 L^2\{(1-a)\sin 2\alpha\}^2 = \|L=1\| = \{0.5(1-a)\sin 2\alpha\}^2 \tag{S5}$$

$$I_{scx} = \left(\frac{E_{scx}}{E_x}\right)^2 = L^2\{(1-a)\cos^2\alpha + a\}^2 = \|L=1\| = \{(1-a)\cos^2\alpha + a\}^2 \tag{S6}$$

Figure S1 shows the variation of the scattered field $x$- and $y$- components as a function of orientation angle of nano-scatterers $\alpha$ with different aspect ratio ($a = w/L$).

(2) SPP scattering by anisotropic nano-scatterers II

We consider the SPP incidence on a prolate spheroid (instead of nano-scatterer of width $w$ and length $L$, $w < L$), inclined with respect to the SPP propagation direction. We further assume that the scattered field can be decomposed into a component along nano-scatterer axis and perpendicular to it, and that these component amplitudes are proportional to the corresponding elements of the polarizability tensor[1]:

$$E_{scx} \sim E_x \cos\varphi\, \alpha_L \cos\varphi + E_x \sin\varphi\, \alpha_w \sin\varphi \sim (\alpha_L - \alpha_w)\cos^2\varphi + \alpha_w \tag{S7}$$

$$E_{scy} \sim E_x \cos\varphi\, \alpha_L \sin\varphi - E_x \sin\varphi\, \alpha_w \cos\varphi \sim 0.5(\alpha_L - \alpha_w)\sin 2\varphi \tag{S8}$$

where

$$\alpha_{w,L} = \frac{\varepsilon_0 \varepsilon_r V_p (\varepsilon_p - \varepsilon_r)}{\varepsilon_r + (\varepsilon_p - \varepsilon_r) m_{w,L}} \tag{S9}$$

$$m_L = \frac{1-\rho^2}{\rho^3}\left(\ln\sqrt{\frac{1+\rho}{1-\rho}} - \rho\right), \left(\rho = \sqrt{1-a^2}, a = \frac{w}{L}\right) \tag{S10}$$

$$m_w = 0.5(1 - m_L) \tag{S11}$$

$$\varepsilon_r = 1, \varepsilon_p \cong 2 \text{ (HSQ)} \tag{S12}$$

$$\alpha_{w,L} \sim \left(\varepsilon_r + (\varepsilon_p - \varepsilon_r) m_{w,L}\right)^{-1} \tag{S13}$$

For the case of $a = \frac{w}{L} \cong 0$, it can be derived that $m_w \cong 0.5$, $m_L \cong 0$, so we get:

$$\alpha_{w,L} \sim \left(\varepsilon_r + (\varepsilon_p - \varepsilon_r) m_{w,L}\right)^{-1} \cong (0.67, 1) \tag{S14}$$

We then consider the SPP incidence on a prolate spheroid (instead of nano-scatterer of width $w$ and length $L$, $w \ll L$), inclined with respect to the radial SPP propagation direction $\theta$ as shown in Fig. S2a. The scattered field of $x$- and $y$- components can be derived as:

$$\begin{aligned} E_{scx} &\sim E\cos(\theta - \varphi)\alpha_L \cos\varphi - E\sin(\theta - \varphi)\alpha_w \sin\varphi \\ &\sim \alpha_L \cos(\theta - \varphi)\cos\varphi - \alpha_w \sin(\theta - \varphi)\sin\varphi \end{aligned} \tag{S15}$$

$$E_{scy} \sim \alpha_L \cos(\theta - \varphi)\sin\varphi + \alpha_w \sin(\theta - \varphi)\cos\varphi \tag{S16}$$

The best angle $\varphi$ that would maximize $E_x$ could be obtained by analytical derivation:

$$\frac{\partial E_{scx}}{\partial \varphi} \sim (\alpha_L - \alpha_w)\{-\cos(\theta - \varphi)\sin\varphi + \sin(\theta - \varphi)\cos\varphi\} = (\alpha_L - \alpha_w)\sin(\theta - 2\varphi) \tag{S18}$$

$$\frac{\partial E_{scx}}{\partial \varphi} = 0 \rightarrow \varphi = \frac{\theta}{2} \tag{S19}$$

The scattered field component $E_x$ reaches its maximum for $\varphi = \theta/2$ for any asymmetry. But we still have to find the best angle $\varphi$ that would maximize $E_x/E_y$ ratio, because this would increase the efficiency of SPP scattering into the wanted field component. It can be expected that this angle will not be far away from that maximizing $E_x$ field component. According to Eq. $S15$ and $S16$, the variation of $E_{scx}$, $E_{scy}$, and $E_{scx}/E_{scy}$ as the function of $\varphi$ at different SPP propagation direction $\theta$ are presented in Fig. S2. It can be observed that the best angle $\varphi$ that would maximize $E_x$, $E_x/E_y$ ratio varies with $\theta$ even though with different amplitude. The best angles $\varphi(\theta)$ that would maximize $|E_{sc}/E_{scy}|$ is presented in Fig. S2f. This calculation of ideal case demonstrates that there are optimal orientations of anisotropic scatters for achieving very high $|E_{scx}/E_{scy}|$.

(3) SPP scattering by realistic nano-scatterers

We should choose realistic nano-scatterer dimensions (height $h$, width $w$ and length L) in our device. We consider the SPP incidence on a nanobrick characterized by two real (positive) elements of the polarizability tensor, $\alpha_L$ and $\alpha_w$, responsible for field scattering of the corresponding polarizations: along the long side $L$ and the short side $w$ of the nanobrick, respectively. The effective mode index of the SPP wave is evaluated as 1.18, then we get $\varepsilon_{reff} = 1.39, \varepsilon_p \cong 2$ (HSQ), $\alpha_w \cong 0.60, \alpha_L \cong 0.68$ according to $Eq.(S10 - S13)$. Similarly, it can be derived that the scattered field component $E_x$ reaches its maximum $\left(\frac{\partial E_{scx}}{\partial \varphi} = 0\right)$ for $\varphi = 0.5\theta$ for any asymmetry. Next, we find the best angle $\varphi$ that would maximize $E_x/E_y$ ratio, which would increase the efficiency of SPP scattering into the wanted field component.

$$E_{scx} = \Delta\alpha\{\cos(\theta - \varphi)\cos\varphi + R\cos\theta\}, \quad (S20)$$

$$E_{scy} = \Delta\alpha\{\cos(\theta - \varphi)\sin\varphi + R\sin\theta\} \quad (S21)$$

where $R = \frac{\alpha_w}{\alpha_L - \alpha_w} = \frac{\alpha_w}{\Delta\alpha}$. Note that $E_{scy}(\theta = 0) = \Delta\alpha\cos\varphi\sin\varphi = 0$ for $\varphi = 0, 0.5\pi$, bu $E_{scx}(\theta = 0) = \Delta\alpha\{(\cos\varphi)^2 + R\} \to$ max for $\varphi = 0$.

$$\Gamma = \frac{E_{scx}}{E_{scy}} = \frac{\cos(\theta - \varphi)\cos\varphi + R\cos\theta}{\cos(\theta - \varphi)\sin\varphi + R\sin\theta} \quad (S22)$$

$$\frac{\partial \Gamma}{\partial \varphi} = \|\theta - \varphi = \alpha\| = -\frac{(\cos\alpha)^2 + R\cos 2\alpha}{(\cos(\theta - \varphi)\sin\varphi + R\sin\theta)^2} = 0 \quad (S23)$$

According to Eq. S23, the relation between $\theta$ and $\varphi$ can be obtained:

$$\pi - 2\alpha = \pm \arccos\left(\frac{\alpha_L - \alpha_w}{\alpha_L + \alpha_w}\right) \to \frac{\pi}{2} - \theta + \varphi = \pm\frac{1}{2}\arccos\left(\frac{\alpha_L - \alpha_w}{\alpha_L + \alpha_w}\right) \quad (S24)$$

Thus, the $E_x/E_y$ ratio reaches its maximum/minimum for $\varphi = \theta - 0.5\pi \pm$ constant depending on the asymmetry. In our cases: $\pm\frac{1}{2}\arccos\left(\frac{\alpha_L-\alpha_w}{\alpha_L+\alpha_w}\right) \cong 43°$ and thus $\varphi = \theta - 47°$. To validate this theoretical prediction, the 3D simulations are conducted by sweeping $\theta$ and $\varphi$ for calculating the outcoupling of a plane SPP wave incident on periodic arrays composed of nanobricks in a rectangular ($7 \times 8$ $\mu m^2$) area (Fig. S3a). The array period is set equal to the SPP wavelength $\lambda_{spp} \cong 510$ nm to ensure the outcoupling in the normal (to the surface) direction. Fig. S3b demonstrated the simulated LP performance $\left(log\left(\frac{I_x}{I_y}\right)\right)$ of outcoupled light as a function of orientation angle ($\varphi$) and azimuthal direction ($\theta$) of propagating SPPs. The relation between optimum angle $\varphi$ and $\theta$ is nearly consistent with the theoretical predictions, following $\varphi \cong \theta - 45°$ for $\theta > 15°$.

Then we can build the array of nanobricks properly aligned in accordance with our optimization procedure. Note that the dependence we use, $\varphi(\theta)$, found for the first quadrant should be mirrored in the nanobrick inclination as shown below. The important part is that left nanobricks should be positioned along the circles that are at different (by $\lambda_{spp}/2$) distances from the QE, compared to the right nanobricks, because they should scatter out-of-phase with respect to each other, compensating the difference in the SPP propagation and therefore the in-plane SPP component. Spacing $d$ between nanobricks along a circle should be chosen $< \lambda_{spp}/2$ and kept constant for different radii. Nanobricks close to $\theta = \pi/2$ will be absent, but a nanobrick at $\theta = 0$ should always be present as shown in Fig. 3 of the main text.

**Supplementary Note 2 | Polarization and topological charge measurement**

The Stokes polarization parameters $[S_0, S_1, S_2, S_3]$ and topological charge of the single photon emission are measured based on setup in Fig. S8. The emitted light pass through a quarter wave plate (retardation angle $\varphi$), then followed by a linear polarizer with its transmission axis aligned at an angle $\theta$ to the $x$ axis. The intensities $I(\theta, \varphi)$ with four different pairs are measured to calculate the four Stokes parameter. The first three Stokes parameter is measured by rotating the polarizer to angle $\theta = 0°$, $90°$, $45°$, and $-45°$ respectively (remove the quarter wave plate). The parameter $S_3$ is measured by the quarter wave retarder ($\theta = \pm 45°, \varphi = 90°$) and linear polarizer ($\theta = 0°$). The Stokes parameter is derived as:

$$s_0 = I(0°, 0°) + I(90°, 0°)$$
$$s_1 = I(0°, 0°) - I(90°, 0°)$$
$$s_2 = I(45°, 0°) - I(-45°, 0°)$$
$$s_3 = I(45°, 90°) - I(-45°, 90°) \quad (S25)$$

in which $s_1$, $s_2$, and $s_3$ are then normalized to the corresponding total intensity ($s_0$) obtained in each measurement. The Stokes parameter $S_1$ of photon emission is calculated as:

$$S_1 = s_1/\sqrt{(s_1)^2 + (s_2)^2 + (s_3)^2} \quad (S26)$$

To characterize the OAM states of LP emission, the topological charge of emission is measured by projecting the far-field emission to different spiral phase distributions of spatial light modulator (SLM). When the topological charge of emission is opposite with the SLM phase, it will form a spot in the far-field. A horizontal linear polarizer is used to decompose the far-field emission to LPx component. Then, the radiated emission pass through SLM and detected by the CMOS in the Fourier plane.

**Supplementary Note 3 | Quantum state tomography**

To characterize the entangled states, quantum state topography is performed based on the polarization and OAM coincidence. The output state of the entangled photon pairs can be written as $|\psi\rangle = a_x|X\rangle|\ell_x = 0\rangle + a_y e^{i\delta}|Y\rangle|\ell_y = -1\rangle)$, $\delta$ is the phase between two entangled states, $a_x$ and $a_y$ are real positive numbers with $a_x^2 + a_y^2 = 1$, as required for normalization. In our cases, $|X\rangle$ and $|Y\rangle$ are considered as the polarization basis, $|\ell = 0\rangle$ and $|\ell = -1\rangle$ are considered as OAM basis. After superpositions of polarization and OAM basis, we get the following states: $|D\rangle = (|X\rangle + |Y\rangle)/\sqrt{2}$, $|R\rangle = (|X\rangle - i|Y\rangle)/\sqrt{2}$, $|+\rangle = (|\ell = 0\rangle + |\ell = -1\rangle)/\sqrt{2}$, $|i\rangle = (|\ell = 0\rangle - i|\ell = -1\rangle)/\sqrt{2}$. Here we show 16 combination ways of polarization and phase to fulfill the topographically complete set of measurements $\hat{P}_{ij} = \hat{\mu}_i \otimes \hat{v}_j^2$.

For the polarization basis:

$$\hat{\mu}_1 = |X\rangle\langle X|$$

$$\hat{\mu}_2 = |Y\rangle\langle Y|$$

$$\hat{\mu}_3 = |D\rangle\langle D| = \left(\frac{|X\rangle + |Y\rangle}{\sqrt{2}}\right)\left(\frac{|X\rangle + |Y\rangle}{\sqrt{2}}\right)^\dagger$$

$$\hat{\mu}_4 = |R\rangle\langle R| = \left(\frac{|X\rangle + |Y\rangle}{\sqrt{2}}\right)\left(\frac{|X\rangle + |Y\rangle}{\sqrt{2}}\right)^\dagger \quad (S27)$$

For the OAM basis:

$$\hat{v}_1 = |\ell = 0\rangle\langle\ell = 0|$$

$$\hat{v}_2 = |\ell = -1\rangle\langle\ell = -1|$$

$$\hat{v}_3 = |+\rangle\langle+| = \left(\frac{|\ell = 0\rangle + |\ell = -1\rangle}{\sqrt{2}}\right)\left(\frac{|\ell = 0\rangle + |\ell = -1\rangle}{\sqrt{2}}\right)^\dagger$$

$$\hat{v}_4 = |i\rangle\langle i| = \left(\frac{|\ell = 0\rangle - i|\ell = -1\rangle}{\sqrt{2}}\right)\left(\frac{|\ell = 0\rangle - i|\ell = -1\rangle}{\sqrt{2}}\right)^\dagger \quad (S28)$$

For the polarization regimes, a combination of half wave plate, linear polarizer, and quarter wave plate is used to decompose the single photon emission to different basis including $X$, $Y$, $D$, $R$ (setup presented in Fig. S8). For phase regimes, the single photon emission pass through SLM with the prepared holograms of the OAM basis, including $|\ell = 0\rangle$, $|\ell = -1\rangle$, $|\ell = 0\rangle + |\ell = -1\rangle$, $|\ell = 0\rangle - i|\ell = -1\rangle$. Then the density matrix can be reconstructed and estimated by Maximum Likelihood method to find physical states with measured intensities by ensemble of all qubit states[3]. The characterizations of entangled polarization-OAM source are performed with the pump power of 210 μW.

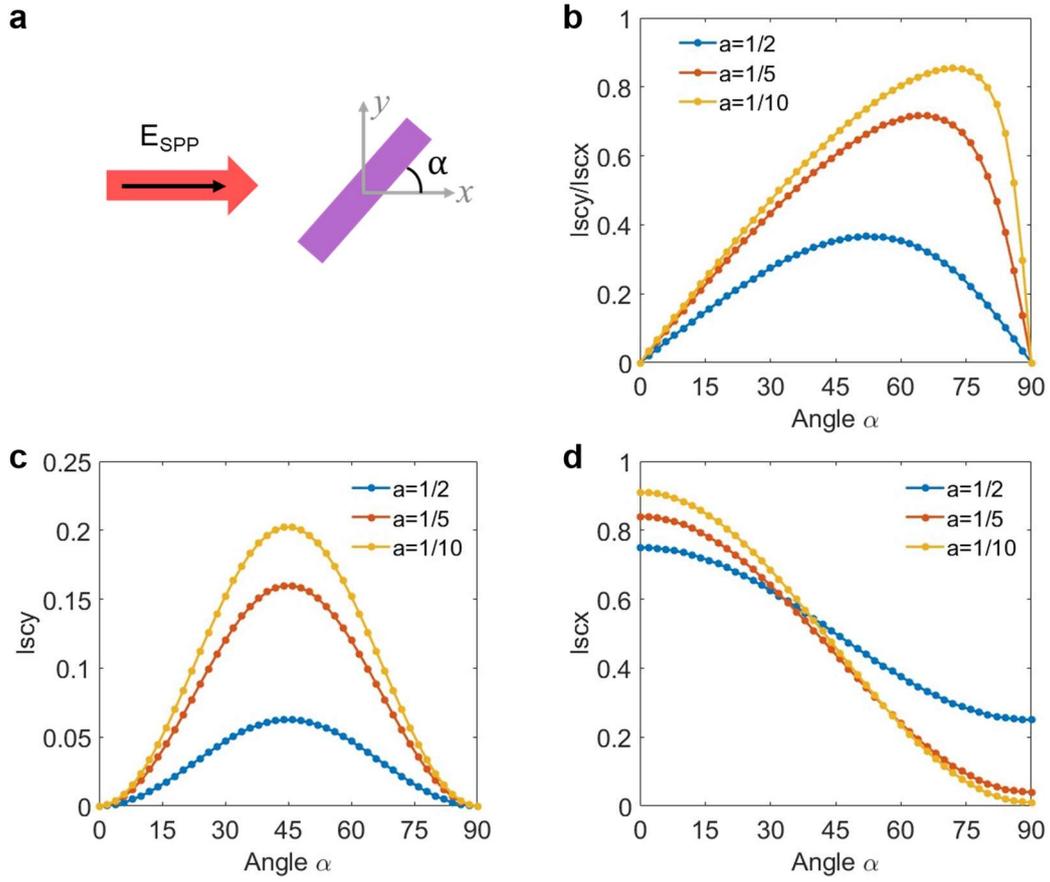

Fig. S1. The scattered field by ideal nano-scatterers with different aspect ratio. a, Normal incident SPP scattered by anisotropic scatterers. b-d, Regimes of the scattering electric fields: Variation of (b) $I_{scy}/I_{scx}$, (c) $I_{scy}$, (d) $I_{scx}$ as a function of orientation angle $\alpha$.

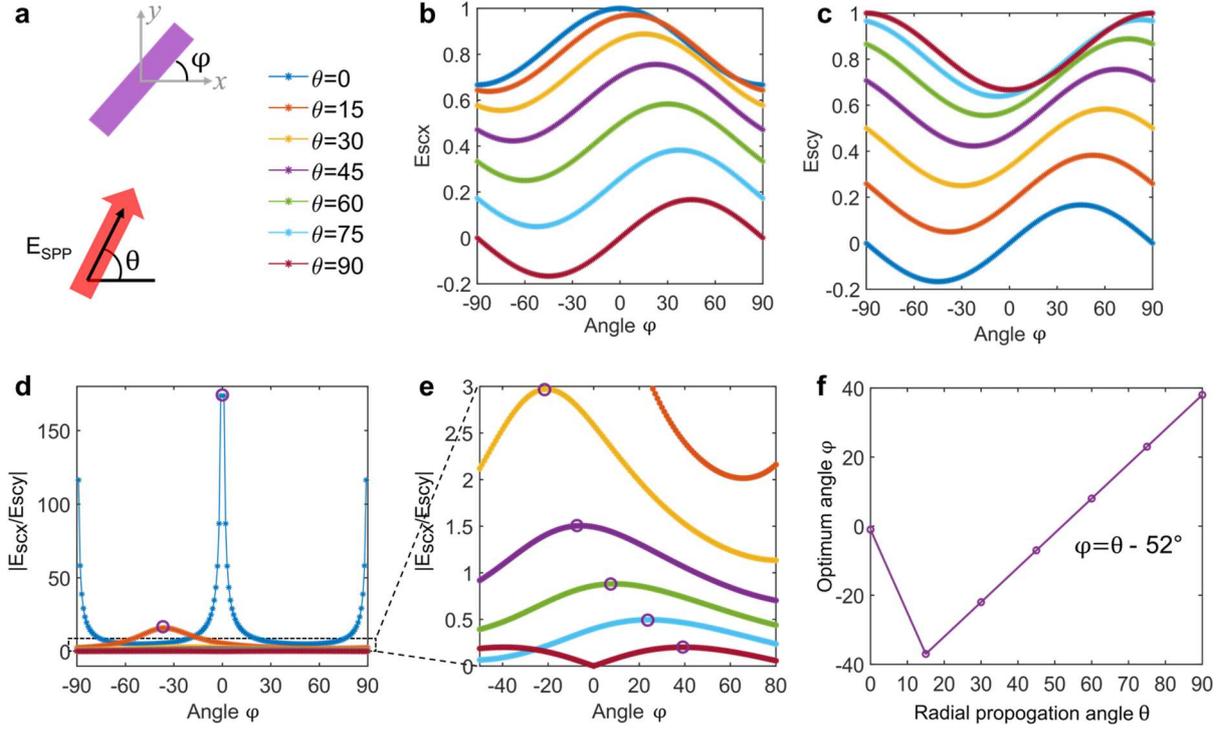

**Fig. S2. Theoretical prediction of the scattered field by a prolate spheroid made of HSQ (ideal case). a**, Schematic of SPP incidence on the anisotropic scatterer with propagation direction $\theta$. **b**, **c**, The scattering electric fields: (**b**) $E_{scy}$ and (**c**) $E_{scx}$ as a function of orientation angle $\varphi$ of anisotropic scatterer. **d-e**, The ratio of scattering electric fields $|E_{scx}/E_{scy}|$ as a function of orientation angle $\varphi$ with different scale range. **f**, The best angles $\varphi(\theta)$ that would maximize $|E_{scx}/E_{scy}|$ ratio.

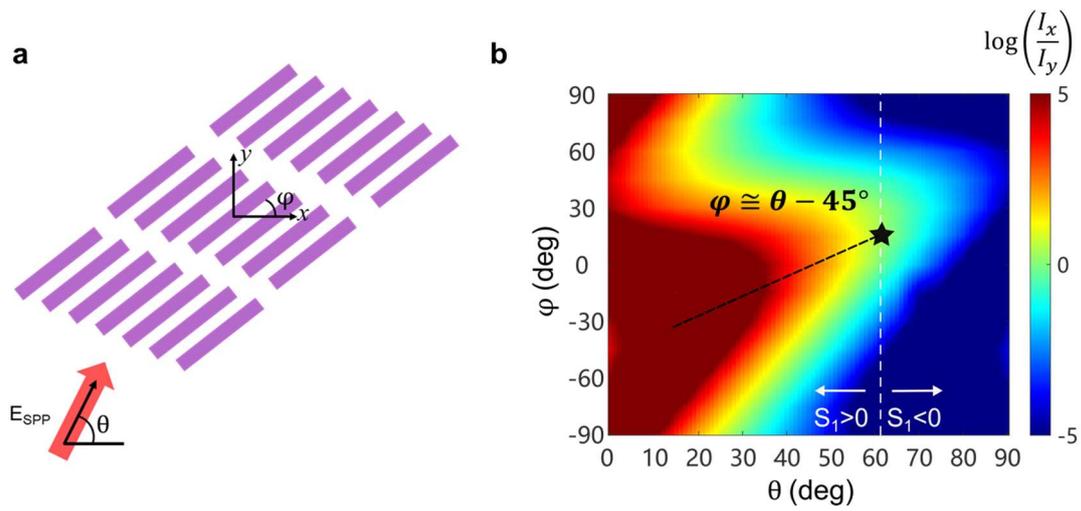

**Fig. S3**. **Simulated relation between optimum angle $\varphi$ and $\theta$. a**, Schematic of a plane SPP wave incident on a rectangular periodically array composed of realistic nanobricks. **b**, Simulated of LP performance as a function of $\theta$ and $\varphi$ (map) and the theoretical predict best angles $\varphi(\theta)$ that would maximize $|E_{scx}/E_{scy}|$ ratio (black line).

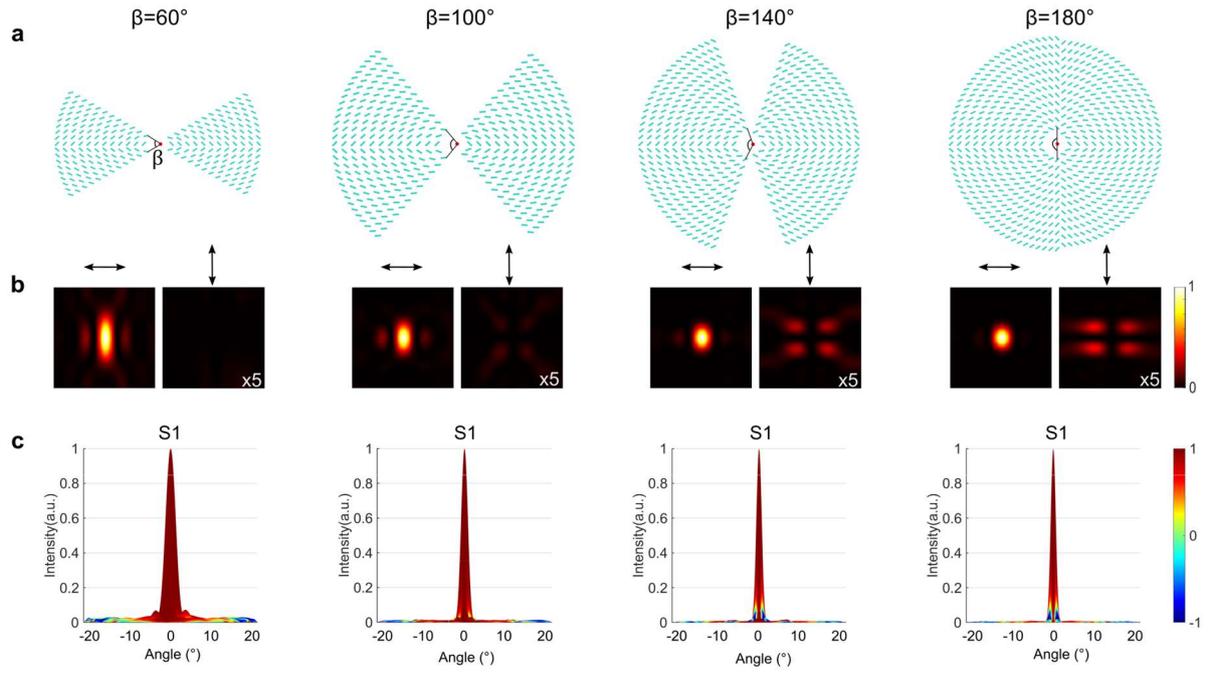

**Fig. S4 Influence of distributed angle $\beta$ on the performance of linear polarized emission. a**, Top view of QE-coupled metasurfaces with different distributed angle ($\beta$ = 60°, 100°, 140°, and 180°). **b**, Far-field intensity distribution of components along *x*- and *y*- axis within numerical aperture NA = 0.2. **c**, The superimposed intensity and degree of linear polarization (the color represents Stokes parameter $S_1$).

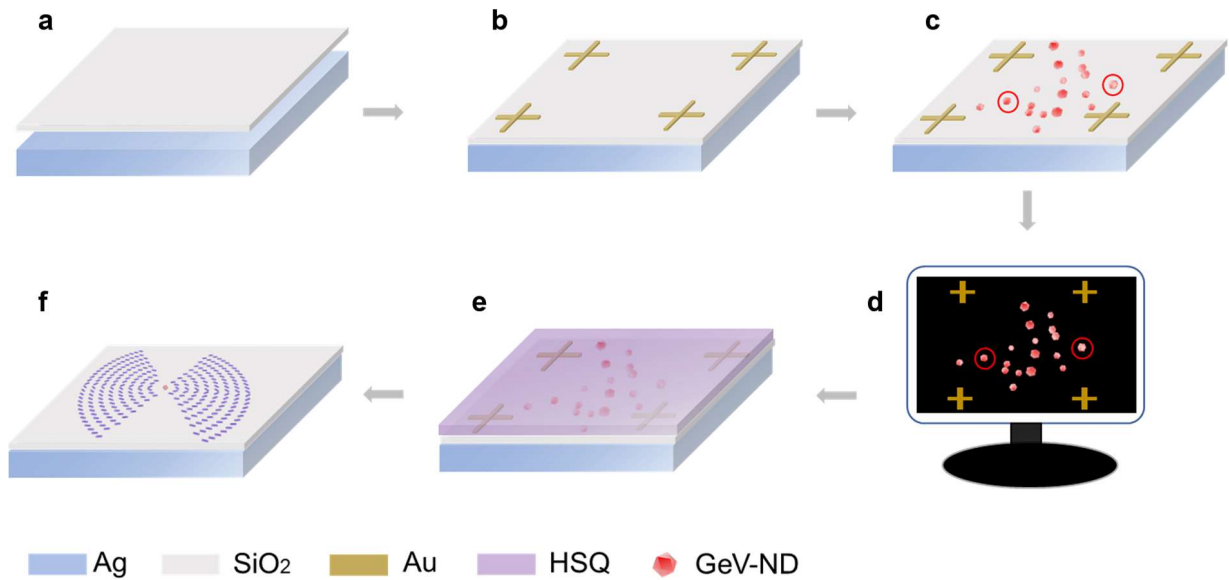

**Fig. S5. Fabrication process of photon sources with single GeV-ND. a**, Deposition of 120 nm Ag and 20 nm SiO$_2$ on the silicon substrate. **b**, The alignment gold markers are fabricated by EBL, gold deposition, and lift-off process. **c**, Spin coating solution of GeV-NDs. **d**, Determine the position of ND-NVs. Single-photon GeV-ND is searched by the fluorescence scan with a radially polarized excitation laser beam (532 nm), the position of which is determined by the microscope image with the help of alignment markers at the corners of a 100×100 $\mu m^2$ area. **e**, Spin coating HSQ and baked at the hotplate to form 150 nm HSQ layer. **f**, Metasurfaces are fabricated around single GeV-ND by EBL.

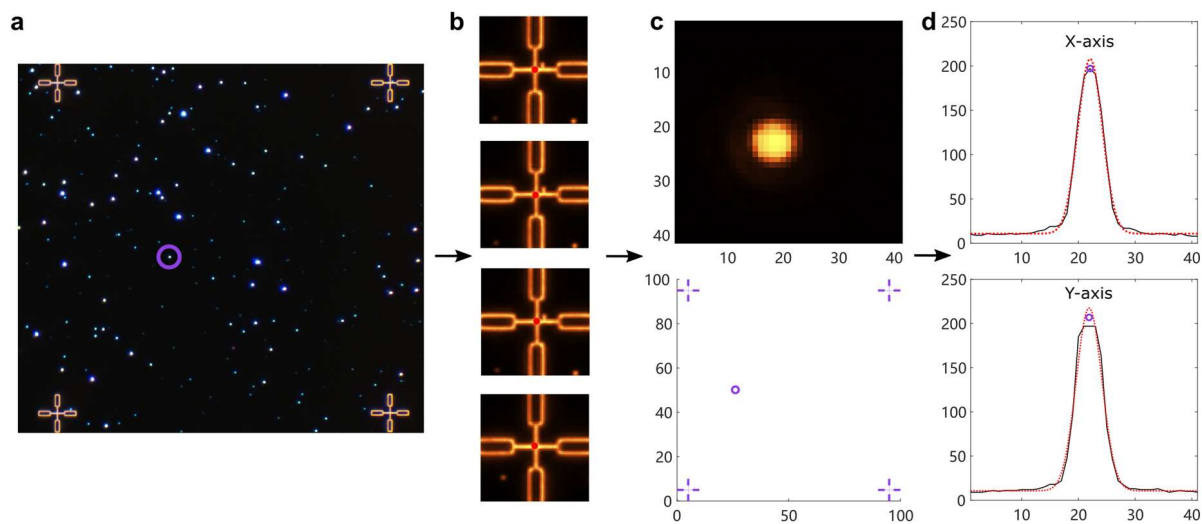

**Fig. S6. Procedure to determine position of nanodiamond. a**, Dark-field microscope image of marker and randomly dispersed GeV-NDs. **b**, Determine the coordinate basis by markers. **c**, Select a nanodiamond. **d**, Determine the position ($x,y$) of the selected nanodiamond with Gaussian fit.

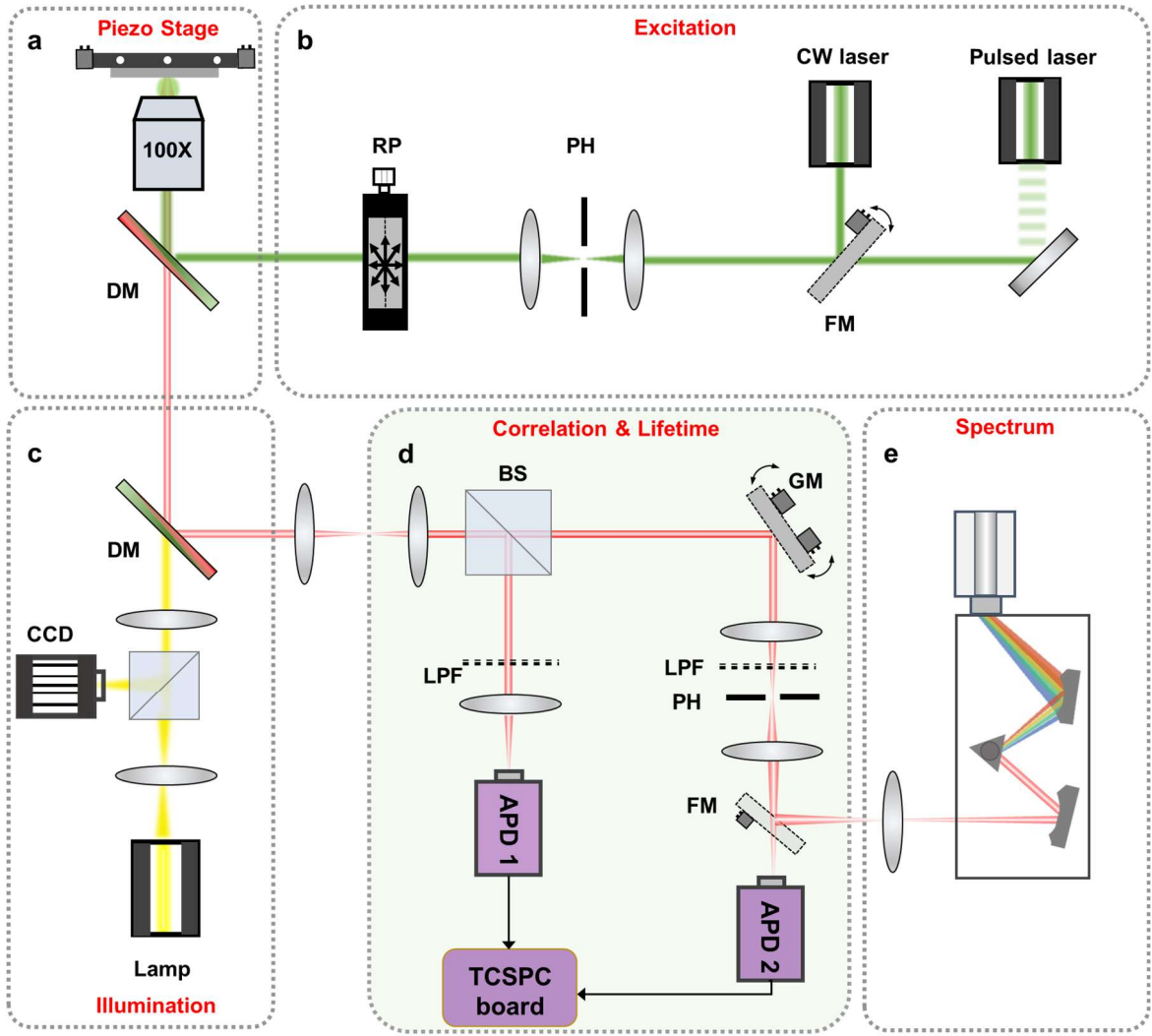

**Fig. S7. Experimental setup for characterizing single photon properties. a**, Sample stage. The piezo-stage allows for locating GeV-NDs when scanning fluorescence maps. **b**, Incident light (532 nm) for exciting the nanodiamond. Pulse laser is used for measuring lifetime. **c**, Illumination part is for finding the fabricated QE-coupled metasurfaces. **d**, Characterization for fluorescence image, correlation, decay-rate. Fluorescence photon rate is recorded by avalanche photo diode (APD1), which is filtered from the laser light by a set of dichroic mirrors (DM) and a long pass filter (LPF). Correlation measurements is recorded by histogramming the timing delay between photon detection events between APD1 and APD2 in a start-stop configuration, using an electronic timing box (Picharp-300; Pico quant). The lifetime data are measured by the time correlated single photon counting (TCSPC) and fitted by the single exponential model ($Ae^{-t/\tau}$ + constant). **e**, Characterization for spectrum. CW: continuous wave, RP: Radial Polarization Converter, PBS: polarized beam splitter, PH: pinhole, DM: dichroic mirror, LP: linear polarization, QWP: quarter-wave plate, LPF: 550 nm long pass filter, FM: flip mirror, GM: galvanometric mirror. APD: Avalanche Photodiode.

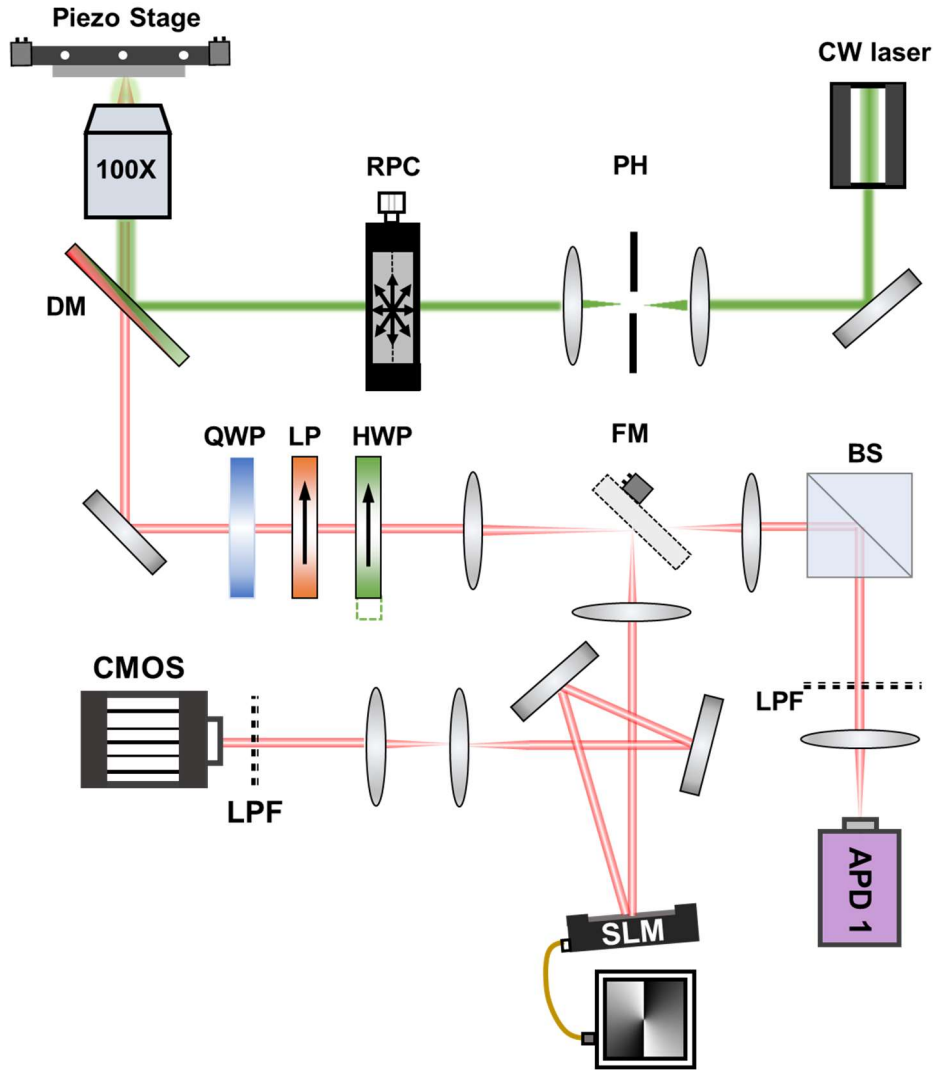

**Fig. S8. Experimental setup for characterizing polarization states and topological charges.** The Stokes parameters were measured to characterize the polarisation. $s_1$ and $s_2$ were measured by orienting the linear polarizer and taking Fourier plane images. $s_3$ was measured by the combination of a quarter-wave plate and a linear polarizer. An SLM was used to generate the computed holograms with different topological charges. The reflected light from the incident light was filtered out by a set of dichroic mirrors (Semrock FF535-SDi01/FF552-Di02) and with a long-pass filter 550 nm (Thorlabs FELH0550) and a band-pass filter 605 nm ± 8 nm.

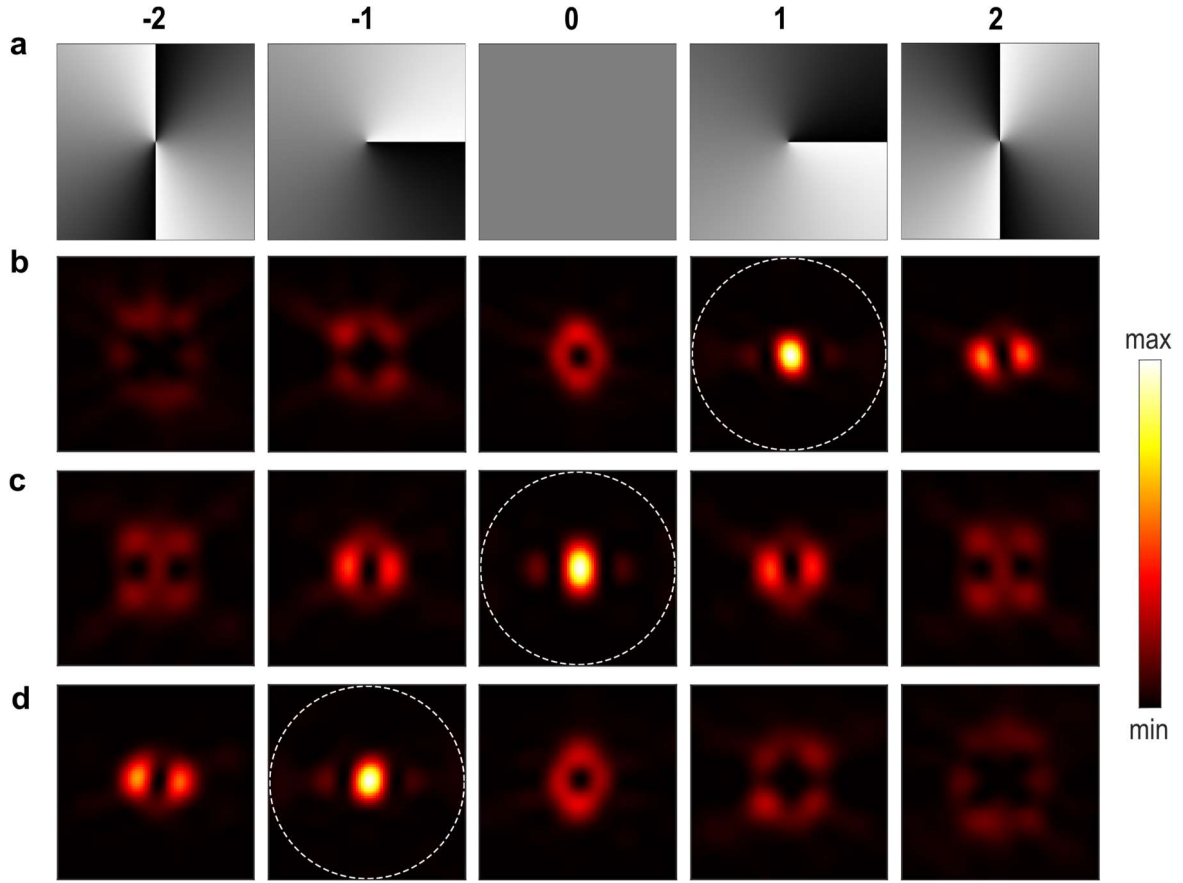

**Fig. S9. Simulated vortex states from LP-OAM single photon source. a**, Phase distribution for holograms with topological charges with -2, -1, 0, 1 and 2. **b-d**, Simulated intensity distributions of the LPx components of the single-photon emissions that are projected to different holograms for characterizing the QE-coupled metasurface shown in Fig. 3, proving the topological charge of $\ell = -1$, 0, and $+1$ from top to bottom. The dashed line indicates the area of NA = 0.2.

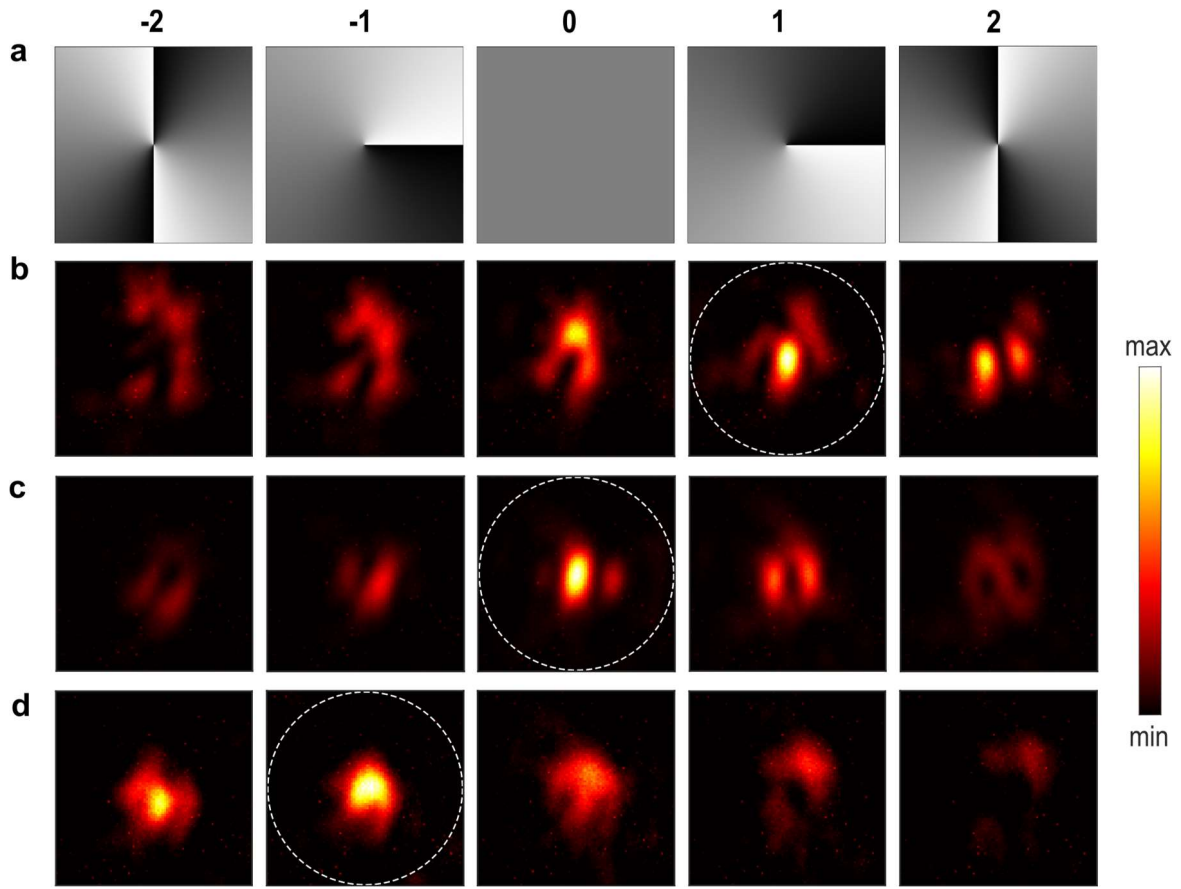

**Fig. S10. Measured vortex states from LP-OAM single photon source. a**, Phase distribution for holograms with topological charges with -2, -1, 0, 1 and 2. **b-d**, Measured intensity distributions of the LPx components of the single-photon emissions that are projected to different holograms for characterizing the QE-coupled metasurface shown in Fig. 3. The dashed line indicates the area of NA = 0.2.

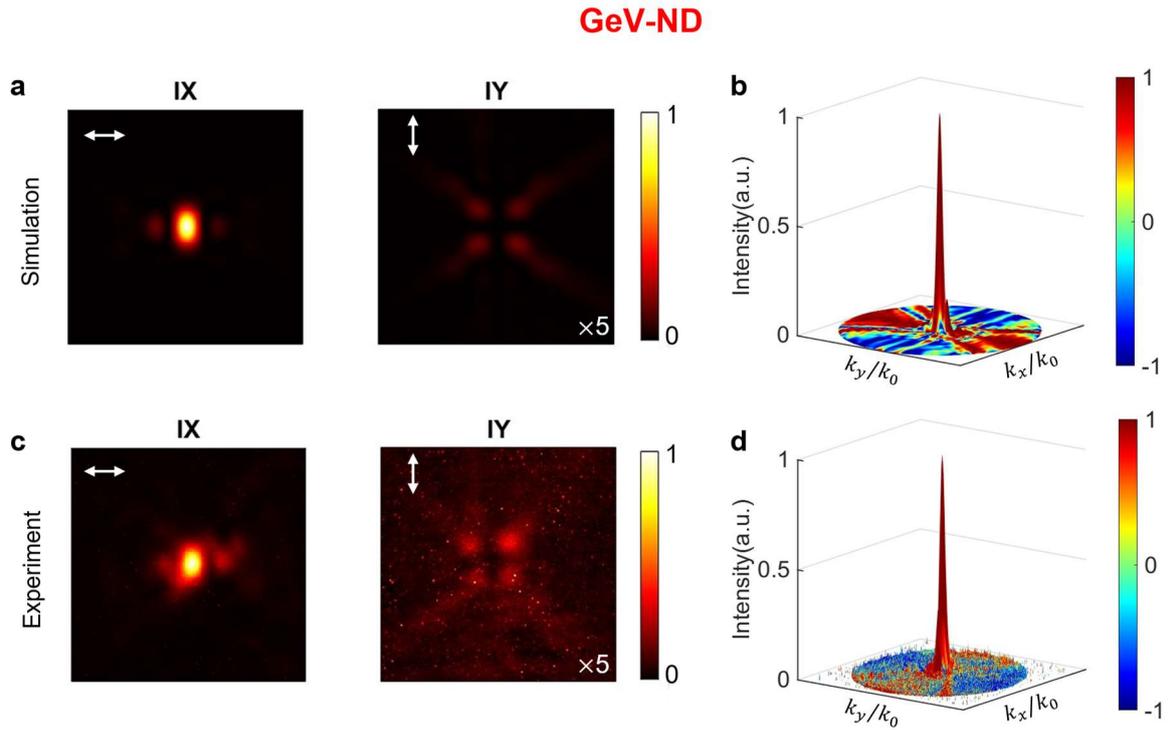

**Fig. S11. Characteristics of linear polarization of GeV-ND single photon source.** The device is the first configuration of Fig. 3A in main text. **a**, Simulated far field intensity distribution of LPx and LPy component. **b**, Simulated 3D representation of the superimposed intensity and Stokes parameter $S_1$ distribution. **c**, Measured far field intensity distribution of LPx and LPy component. **d**, Measured 3D representation of the superimposed intensity and Stokes parameter $S_1$ distribution.

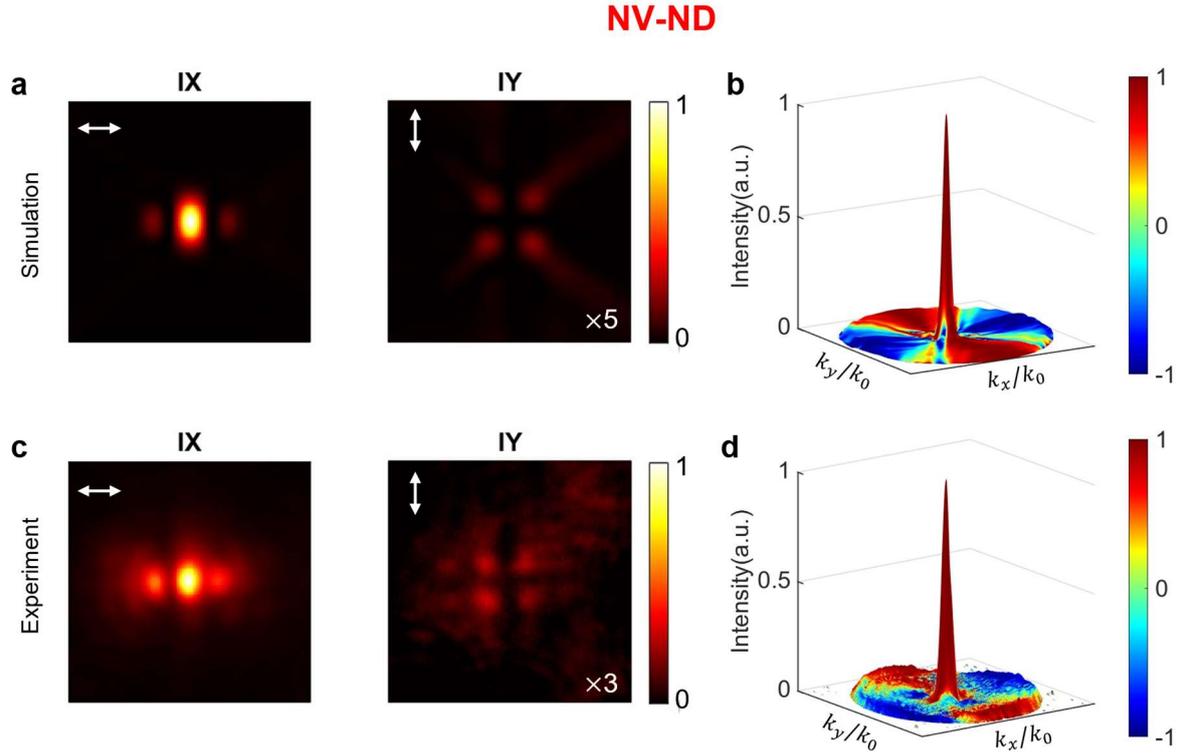

**Fig. S12. Characteristics of the linear polarized states of NV-ND multiple photon source. a**, Simulated far field intensity distribution of LPx and LPy component. **b**, Simulated 3D representation of the superimposed intensity and Stokes parameter $S_1$ distribution. **c**, Measured far field intensity distribution of LPx and LPy component. **d**, Measured 3D representation of the superimposed intensity and Stokes parameter $S_1$ distribution.

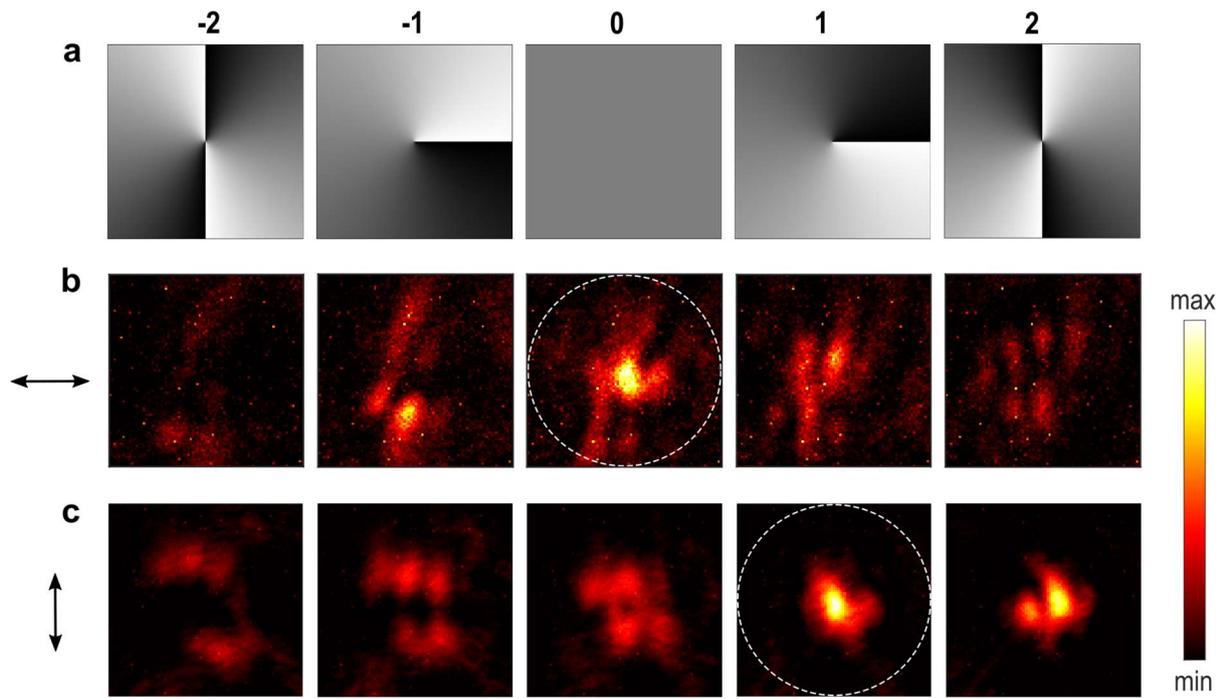

**Fig. S13. Measured vortex states from entangled LP-OAM single photon source.** Intensity distributions of LPx and LPy components of the single-photon emissions that are projected to different holograms, which verifies the two LP channel carries different topological charge of 0 and -1.

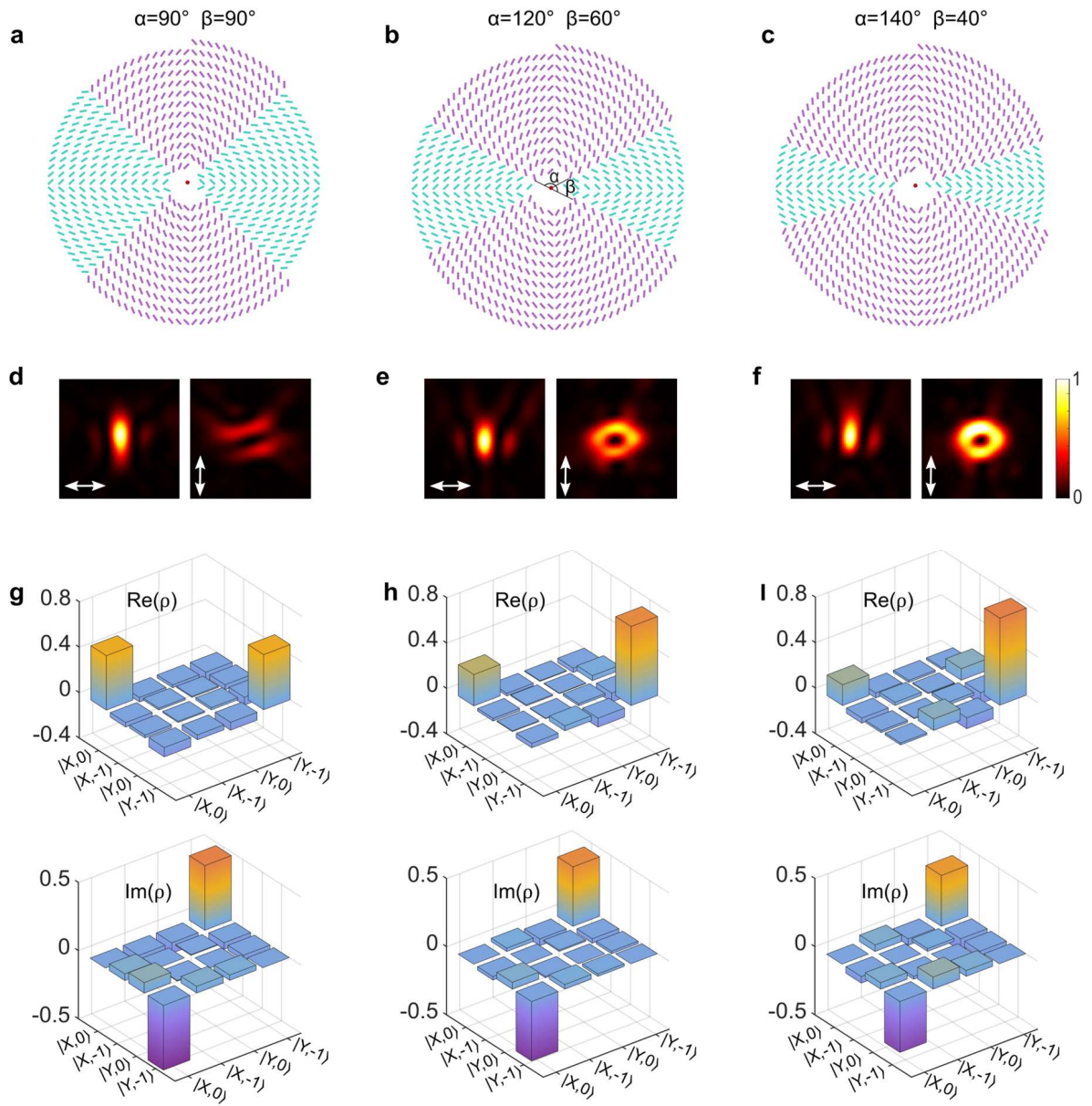

**Fig. S14. Influence of distributed angles on the performance of entangled LP-OAM emission. a-c**, Top view of QE-coupled metasurfaces with different distributed angles. **d-f**, Far-field intensity distribution of components along *x*- and *y*- axis within numerical aperture NA = 0.2. **g-i**, The density matrices with real and imaginary parts, demonstrating that the entangled state could be modulated by controlling the angles of two parts of metasurfaces.